\documentclass[runningheads]{llncs}
\usepackage{graphicx}
\usepackage{cite}
\usepackage{siunitx}
\usepackage{mathtools}
\usepackage[misc]{ifsym}

\begin{document}
\title{Uncertainty Estimation in Medical Image Localization: Towards Robust Anterior Thalamus Targeting for Deep Brain Stimulation}
\titlerunning{Uncertainty Estimation for ANT-DBS Localization}
%
\author{Han Liu\inst{1}\Letter\and
Can Cui\inst{1} \and
Dario J. Englot\inst{2} \and
Benoit M. Dawant\inst{1}}
\authorrunning{H. Liu et al.}
%
\institute{Department of Electrical Engineering and Computer Science, Vanderbilt University, Nashville, TN 37235, USA\and 
Department of Neurosurgery, Vanderbilt University Medical Center, Nashville, TN, 37235, USA\\
\email{han.liu@vanderbilt.edu}}

\maketitle              
\begin{abstract}
Atlas-based methods are the standard approaches for automatic targeting of the Anterior Nucleus of the Thalamus (ANT) for Deep Brain Stimulation (DBS), but these are known to lack robustness when anatomic differences between atlases and subjects are large. To improve the localization robustness, we propose a novel two-stage deep learning (DL) framework, where the first stage identifies and crops the thalamus regions from the whole brain MRI and the second stage performs per-voxel regression on the cropped volume to localize the targets at the finest resolution scale. To address the issue of data scarcity, we train the models with the pseudo labels which are created based on the available labeled data using multi-atlas registration. To assess the performance of the proposed framework, we validate two sampling-based uncertainty estimation techniques namely Monte Carlo Dropout (MCDO) and Test-Time Augmentation (TTA) on the second-stage localization network. Moreover, we propose a novel uncertainty estimation metric called maximum activation dispersion (MAD) to estimate the image-wise uncertainty for localization tasks. Our results show that the proposed method achieved more robust localization performance than the traditional multi-atlas method and TTA could further improve the robustness. Moreover, the epistemic and hybrid uncertainty estimated by MAD could be used to detect the unreliable localizations and the magnitude of the uncertainty estimated by MAD could reflect the degree of unreliability for the rejected predictions.

 \and Pseudo Labels
\end{abstract}
\section{Introduction}
Epilepsy is one of the most common chronic neurological disorders characterized by spontaneous recurrent seizures and affects around 70 million patients worldwide\cite{ref1}. Over 30\% of the epilepsy patients have refractory seizures which may carry risks of structural damage to the brain and nervous system, comorbidities, and increased mortality\cite{ref2}. Deep Brain Stimulation (DBS) is a recently FDA-approved neurostimulation therapy that can effectively reduce the occurrences of refractory seizures by delivering electric impulses to a deep brain structure called the anterior nucleus of the thalamus (ANT). Accurate localization of the ANT target is however difficult because of the well documented variability in the ANT size and shape and thalamic atrophy caused by persistent epileptic seizures \cite{ref3}. Currently, the standard approach to automate this process is the atlas-based technique. While popular, atlas-based methods are known to lack robustness when anatomic differences between atlases and subjects are large. This is particularly acute for ANT-DBS targets that are close to the ventricles, which can be severely enlarged in some patients. 

Over the past decade, DL-based techniques such as convolutional neural networks (CNN) have emerged as powerful tools and have achieved unprecedented performances in many medical imaging tasks. However, to train sufficiently robust and accurate models, deep learning methods typically require large amounts of labeled data, which is expensive to collect, especially in the medical domain. In the case of data scarcity and noisy labels, insufficiently trained models may fail catastrophically without any indication. Hence, it is extremely desirable for deep learning models to estimate the uncertainties regarding their outputs in these scenarios. The predictive uncertainty of neural networks can be categorized into two types: \textit{epistemic} uncertainty and \textit{aleatoric} uncertainty. Epistemic uncertainty, also known as model uncertainty, accounts for the uncertainty in the model and can be explained away by observing more training data. On the other hand, the aleatoric uncertainty is the input-dependent uncertainty that captures the noise and randomness inherent in observations. Recently, uncertainty estimation has also received increasing attention in medical image analysis. Ayhan et al.\cite{ref4} proposed to estimate the heteroscedastic aleatoric uncertainty using TTA for classification task. Nair et al.\cite{ref5} explored the uncertainty estimation for lesion detection and segmentation tasks based on MCDO. Wang et al.\cite{ref6} proposed a theoretical formulation of TTA and demonstrated its effectiveness in uncertainty estimation for segmentation task. Nevertheless, uncertainty estimation for localization tasks has not been well studied.

In this work, we developed a novel two-stage deep learning framework aiming at robustly localizing the ANT targets. To the best of our knowledge, this is the first work to develop a learning-based approach for this task. To overcome the problem of data scarcity, we train the models with the pseudo labels which are created based on the available gold-standard annotations using multi-atlas registration. Moreover, we validate two sampling-based uncertainty estimation techniques to assess the localization performance of the developed method. We also propose a novel metric called MAD for sampling-based uncertainty estimation methods in localization tasks. Our experimental results show that the proposed method achieved more robust localization performance than the traditional multi-atlas method and TTA could further improve the robustness. Lastly, we show that the epistemic and hybrid uncertainty estimated by MAD can be used to detect the unreliable localizations and the magnitude of MAD can reflect the degree of unreliability when the predictions are rejected.

\section{Materials and Methods}
\subsection{Data}
Our own dataset consists of 230 T1-weighted MRI scans from a database of patients who underwent a DBS implantation for movement disorders, i.e., Parkinson Disease or Essential Tremor at Vanderbilt University. The resolution of the images varies from 0.4356$\times$0.4356$\times$1 mm\textsuperscript{3} to 1$\times$1$\times$6 mm\textsuperscript{3}. The ground truth was manually annotated on a different dataset collected by an experienced neurosurgeon. In this dataset, the 3D coordinates of eight ANT targets (one on each side) on four MRI scans were identified and the thalamus mask on one of these volumes was delineated. With the available annotations, we generated the pseudo labels for the ANT targets and for the thalamus masks using multi and single-atlas registration\cite{ref7}. In this study, 200 MRI scans were randomly selected for training and validation and the remaining 30 images were used for testing. For preprocessing, we use trilinear interpolation to resample all the images to isotropic voxel sizes of 1$\times$1$\times$1 mm\textsuperscript{3} and rescale the image intensities to $[0, 1]$.

\subsection{Proposed Method}
Typically, an MRI scan with original resolution cannot be fed to a 3D CNN directly due to the limitation of computational resource. A common approach to solve this problem, i.e., using downsampled images, is not appropriate here because the downsampling operation unavoidably leads to a loss in image resolution. This is a concern in our application because even a few-voxel shift in the deep brain can lead to target predictions that are unacceptable for clinical use. To address this issue, we propose a two-stage framework where the first stage coarsely identifies and crops the thalamus regions from the whole brain MRI and the second stage performs per-voxel regression on the cropped volume to localize the targets at the finest resolution scale. 

\begin{figure}
\includegraphics[width=1\columnwidth]{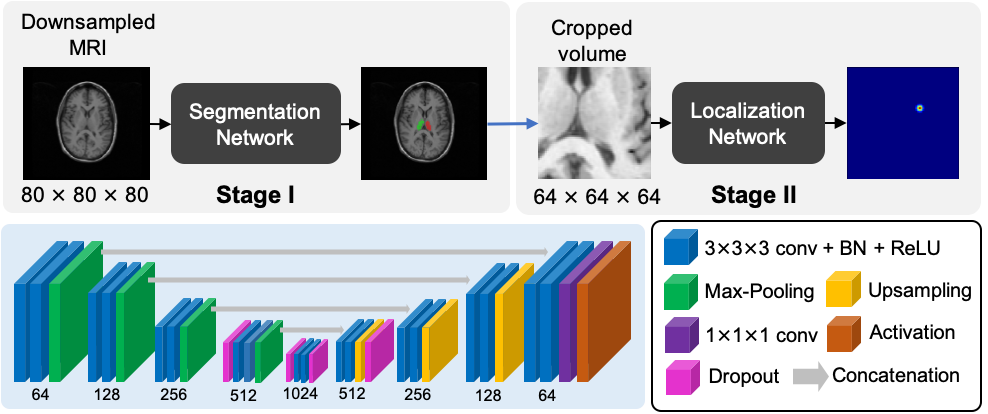}
\centering
\caption{The workflow of the proposed two-stage framework and the 3D U-net architecture we used. The segmentation and localization network share the same architecture. The number below each encode/decoder unit is the channel number of the 3$\times$3$\times$3 convolution kernels.} \label{fig1}
\end{figure}

The workflow of the proposed framework is shown in Figure 1. In this first stage, we train a 3D U-net \cite{ref8} using the downsampled 80$\times$80$\times$80 MRI scans to coarsely segment the thalamus. The output layer of this network has three channels corresponding to background, left thalamus and right thalamus respectively. Once we obtain the segmentation results, we post-process the binary segmentation from each foreground channel by isolating the largest connected component and resample the results back to the original resolution. Thereafter, we compute the bounding box of each isolated component and crop a 64$\times$64$\times$64 mm\textsuperscript{3} volume
around its center. The cropped volume fully encloses the entire left or right thalamus as well as its contextual information. Before passing the volumes to the second stage, we flip the left-thalamus volumes in the left-right direction so that the inputs of the second stage have consistent orientations. In the second stage, we employ another 3D U-net with the same architecture to localize the ANT target by performing per-voxel regression on the cropped volumes. Since the cropped volumes have the same resolution as the original MRI, there is no performance degradation in localization due to loss in resolution. To allow volume-to-volume mapping, we design the ground truth map to be a 3D Gaussian function centered at the pseudo label position with a standard deviation of 1.5 mm. The maximum value is scaled to 1 and any value below 0.05 is set to 0. During the testing phase, the left-thalamus localization maps are flipped back to the original orientation and the voxel with the \textbf{maximum} activation in each localization map is taken as the final prediction. 

\subsection{Uncertainty Estimation}
\subsubsection{Epistemic Uncertainty}
We estimate the epistemic uncertainty of the localization task using the dropout variational inference. Specifically, we train the model with dropout (same as the baseline method) and during the testing phase we perform T stochastic forward passes with dropout to generate Monte Carlo samples from the approximate posterior. Let $y = f(x)$ be the network mapping from input $x$ to output $y$. Let $T$ be the number of Monte Carlo samples and $\hat{W}_{t}$ be the sampled model weights from MCDO. For regression tasks, the final prediction and the epistemic uncertainty can be estimated by calculating the predictive mean $E(y)\approx\frac{1}{T}\sum_{t=1}^{T}f^{\hat{W}_{t}}(x)$ and variance $Var(y)\approx\frac{1}{T}\sum_{t=1}^{T}f^{\hat{W}_{t}}(x)^{T} f^{\hat{W}_{t}}(x)-E(y)^{T}E(y)$ from these samples. 

\subsubsection{Aleatoric Uncertainty}
To estimate the aleatoric uncertainty, we use the TTA technique which is a simple yet effective approach to study locality of testing samples. Recently, Wang et al.\cite{ref6} provided a theoretical formulation for using TTA to estimate a distribution of prediction by Monte Carlo simulation with prior distributions of image transformation and noise parameters in an image acquisition model. In our image acquisition model, we extend this idea by incorporating both spatial transformations $T_{s}$ and intensity transformation ${T_{i}}$  to simulate the variations of spatial orientations and image brightness and contrast respectively. Our image acquisition model can thus be expressed as: $x = T_s(T_i(x_0))$, where $x$ is our observed testing image and $x_0$ is the image without transformations in latent space. During the testing phase, we aim to reduce the bias caused by transformations in $x$ by leveraging the latent variable $x_0$. Given the prior distributions of the transformation parameters in $T_{s}$ and $T_{i}$, we can estimate $y$ by generating N Monte Carlo samples and the $n_{th}$ Monte Carlo sample can be inferred as: $y_n=T_{s_n}(y_{0_n})=T_{s_n}(f(x_{0_n}))=T_{s_n}(f(T_{i_n}^{-1}(T_{s_n}^{-1}(x_n))))$. The final prediction and aleatoric uncertainty can be obtained by computing the mean and variance from the Monte Carlo samples. 

\subsubsection{Maximum Activation Dispersion}
For regression tasks, the uncertainty maps are typically obtained by computing the voxel-wise variance from the Monte Carlo samples. However, this approach fails to generate useful uncertainty maps in our application. In our ground truth maps, the non-zero elements, i.e., foreground voxels within the Gaussian ball, are very sparse compared to the zero elements, and thus more difficult to localize and more likely to produce larger predictive variance. As a result, such uncertainty maps would display higher uncertainty at the predicted targets even if the targets are correctly localized (Figure 2), and thus are not effective for uncertainty estimation regarding the localization performance. To address this issue, we propose a novel metric called Maximum Activation Dispersion (\textbf{MAD}) which can be directly applied to any sampling-based uncertainty estimation technique. This metric measures the consistency of the maximum activation positions of the Monte Carlo samples and ignores the activation variance at the same position. Note that MAD aims at estimating the image-wise uncertainty regarding the overall localization performance instead of voxel-wise uncertainty produced by uncertainty maps. Let N be the number of Monte Carlo samples and $\mathbf{p}_n = (x_{n}, y_{n}, z_{n})$ be the maximum activation position of the $n_{th}$ Monte Carlo sample. The maximum activation dispersion is computed as $\frac{1}{N}\sum_{n=1}^{N}\|{\mathbf{p}_n-\bar{\mathbf{p}}\|}$, where $\|\cdot\|$ is the $L_2$ norm and $\bar{\mathbf{p}}= \frac{1}{N}\sum_{n=1}^{N}\mathbf{p_n}$ is the geometric center of all maximum activation positions.

\begin{figure}
\includegraphics[width=1\columnwidth]{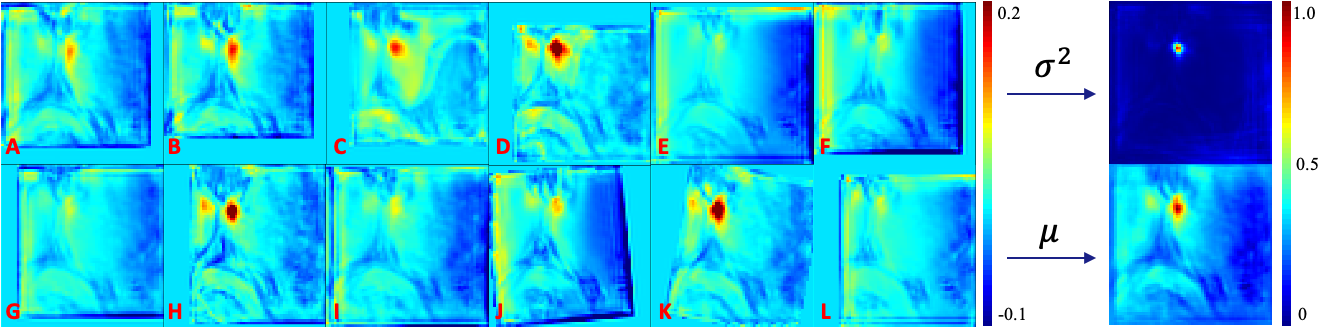}
\centering
\caption{Visualization of some Monte Carlo samples (A-L) by TTA, their mean (localization map) and variance (uncertainty map). The uncertainty map displays higher uncertainties at the correctly localized position and thus is not effective for localization uncertainty estimation.} \label{fig2}
\end{figure} 

\subsection{Implementation Details}
Two five-level 3D U-nets with the same architecture were used in the proposed two-stage framework (Figure 1). In the first stage (segmentation), optimization was performed using the Adam optimizer with a learning rate of \SI{5e-4}, with Dice loss as the loss function, a batch size of 3, and early stopping based on validation loss with patience of 10 epochs. In the second stage (localization), dropout layers were added to allow MCDO. As suggested by Kendall et al.\cite{ref9}, applying dropout layers to all the encoders and decoders is too strong a regularizer. To avoid poor training fit, we followed the best dropout configuration in \cite{ref9} by adding dropout layers with a dropout rate of $p=0.5$ only at the deepest half of encoders and decoders. During training, optimization was performed using the Adam optimizer with a learning rate of \SI{2e-4}, with a batch size of 6, and early stopping based on validation loss with patience of 5 epochs. A weight decay of \SI{5e-4} was used. The weighted mean squared error (WMSE) was used as the loss function to alleviate the class imbalance issue by assigning higher weights to the sparse non-zero entries. The models with the smallest validation losses were selected for final evaluation. 

During the testing phase, we forward passed the testing image once to the deterministic network with dropout turned off (\textbf{baseline}). With a given prior distribution of transformation parameters in image acquisition model, we forward passed the stochastically transformed testing image $N=100$ times to the deterministic network with dropout turned off and transformed the predictions back to the original orientation. The mean and variance (aleatoric uncertainty) of the Monte Carlo samples were obtained (\textbf{baseline + TTA}). In the image acquisition model, the spatial transformations were modeled by translation and rotation along arbitrary axis. The intensity transformation was modeled by a smooth and monotonous function called Bézier Curve, which is generated using two end points $P_0$ and $P_3$ and two control points $P_1$ and $P_2$: $B(t) = (1-t)^3P_0 + 3(1-t)^2tP_1+3(1-t)t^2P_2+t^3P_3, t\in[0,1]$. In particular, we set $P_0 = (0, 0)$ and $P_3 = (1, 1)$ to obtain an increasing function to avoid invalid transformations.
The prior distributions of the spatial and intensity transformation parameters were modeled by uniform distribution $U$ as $s \sim U(s_0, s_1)$, $r \sim U(r_0, r_1)$ and $t \sim U(t_0, t_1)$, where $s$, $r$ and $t$ represent translation (voxels), rotation angle (degrees) and the fractional value for Bézier Curve. In our experiment, we set $s_0 = -10$, $s_1 = 10$, $r_0 = -20$, $r_1 = 20$, $t_0 = 0$ and $t_1 = 1$. Lastly, we forward passed the same testing image to the stochastic network $T=100$ times with dropout turned on with a rate of $p=0.5$ and obtained the mean and variance (epistemic uncertainty) of the Monte Carlo samples (\textbf{baseline + MCDO}).

In our experiment, we evaluated the localization performances of the multi-atlas, baseline, baseline + TTA and baseline + MCDO methods on 30 testing images (60 targets). The axial, sagittal and coronal views centered at the targets predicted by each method were provided to an experienced neurosurgeon to evaluate whether the predicted targets are acceptable for clinical use (the order was shuffled and the evaluator was blind to the method used to predict the target). When the predictions were evaluated as rejected, the evaluator was asked to provide the reasons for rejection. Furthermore, we analyze the aleatoric, epistemic and hybrid (aleatoric + epistemic) uncertainties estimated by MAD on the baseline rejected cases.

\section{Experimental Results}
Our results show that among a total number of 60 targets, 53, 55, 57 and 55 targets were evaluated as acceptable for the multi-atlas, baseline, baseline + TTA and baseline + MCDO respectively. In Figure 3, we show the boxplots of aleatoric, epistemic and hybrid uncertainties estimated by MAD. It can be observed that when the rejected predictions are far away from the acceptable positions (red and blue), their estimated uncertainty correspond to the outliers above the upper whisker in the boxplots of epistemic and hybrid uncertainty. On the other hand, when the rejected predictions are close to the acceptable positions (cyan, green and magenta), their uncertainties fall in the range of upper quartile and the upper whisker (cyan and green) and the range of lower quartile and median (magenta), corresponding to their degree of unreliability. 

\begin{figure}
\includegraphics[width=1\columnwidth]{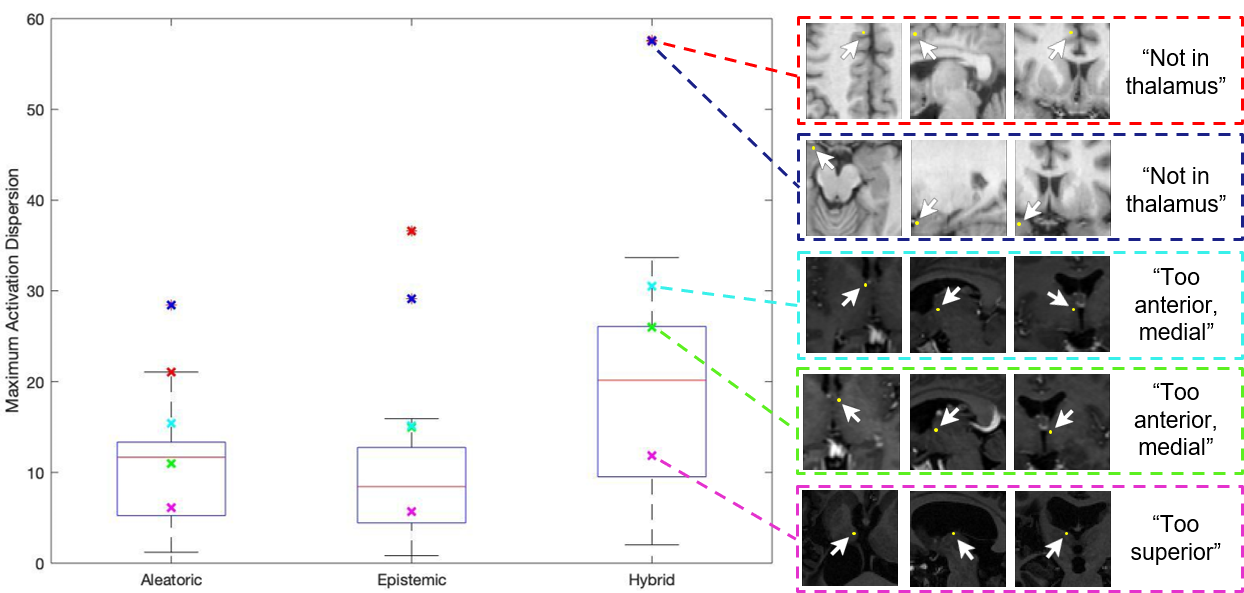}
\centering
\caption{Boxplots of aleatoric, epistemic and hybrid uncertainties estimated by MAD on the testing set (60 targets). The rejected cases of the baseline method (5 cases) are shown in color. The axial, sagittal and coronal views of the rejected targets are shown with the reasons for rejection provided by the evaluator.} \label{fig3}
\end{figure}

It can be observed that the epistemic and hybrid uncertainty estimated by MAD could be used to detect unreliable localizations, i.e., the ones that not even in thalamus. Moreover, the magnitudes of MAD could reflect the degree of unreliability when the predictions were rejected. We also observe that even though the MCDO did not improve the localization robustness compared to the baseline method, the epistemic uncertainty obtained by this technique has great value for detecting the unreliable localizations, i.e., the outliers in the boxplot.

\section{Conclusion}
In this study, we present a two-stage deep learning framework to robustly localize the ANT-DBS targets in MRI scans. Results show that the proposed method achieved more robust localization performance than the traditional multi-atlas method and TTA-based aleatoric uncertainty estimation can further improve the localization robustness. We also show that the proposed MAD is a more effective uncertainty estimation metric for localization tasks.

%
%
%
%

\end{document}